July 30, 2014

**The degenerative evolution from multicellularity to unicellularity during cancer**

Han Chen, Fangqin Lin and Xionglei He

State Key Laboratory of Biocontrol, School of Life Sciences, Sun Yat-sen University, Guangzhou, 510275, China

Correspondence to:
Xionglei He, PhD (University of Michigan)
Room 601, Danqing Hall
South Campus
Sun Yat-sen University
Guangzhou, 510275
China
Tel: 86-20-84110775
Email: hexiongl@mail.sysu.edu.cn

**Abstract**

Theoretical reasoning suggests that human cancer may result from knocking down the genetic constraints evolved for maintenance of the metazoan multicellularity, which, however, requires a critical test.  Using xenograft-based experimental evolution we characterized for the first time the full life history from initiation to metastasis of a tumor at the genomic and transcriptomic levels, and observed metastasis-driving positive selection for generally loss-of-function mutations on a set of multicellularity-related genes, which is further supported by large-scale exome data of clinical tumor samples.  Subsequent expression analysis revealed mainly expression down-regulation of multicellularity-related genes, which form an evolving expression profile approaching that of embryonic stem cells, the cell type with the most characteristics of unicellular life.  The theoretical conjecture predicts that genes born at the emergence of metazoan multicellularity tend to be cancer drivers, which we validated using a rigorous phylostratigraphy analysis on the birth rate of genes annotated by Cancer Gene Census.  Also, the number of loss-of-function tumor suppressors often predominates over activated oncogenes in a typical tumor of human patients.  These data collectively suggest that, different from typical organismal evolution in which gain of new genes/functions is the mainstream, cancer represents a loss-of-function-driven degenerative evolution back to the unicellular "ground state".  This cancer evolution model may explain the enormous inter-/intra-tumoral genetic heterogeneity in the clinic, underlie how distant-organ metastases originate in primary tumors despite distinct environmental requirements, and hold implications for designing effective cancer therapy.

**Introduction**

Unicellular life appeared earlier on the earth; group selection for the public good promoted cooperation of the single-celled individuals, eventually leading to multicellular life [1]. Complex multicellular organisms, including humans, must possess sophisticated genetic constraints that suppress the fitness of individual cells in order to ensure the fitness of the whole organism [2]. However, accidental events such as somatic mutations or viral infections can wipe out such constraints and reactivate the cell's otherwise dormant capacity of seeking for its own fitness, often resulting in cancer [3-5]. In this regard, cancer represents a reversal of the macroevolution from unicellular life to multicellular life [6-8]. This reasonable conjecture has important implications for cancer research, cancer prevention and cancer therapy, but has never been tested rigorously.

We carried out an experimental evolution [9,10] of a human breast cell-derived xenograft tumor in mice to characterize the complete evolutionary history of a tumor, attempting to address the question. Although the micro-environment of mouse mammary gland is not ideal for studying human breast cancer, such a strategy has long been used to study the genetics underlying tumorigenesis, tumor progression and metastasis [11], and has been proven highly successful in human cancer biology [12]. Similar to a pioneer study [13], we engineered a mutated version of human oncogene $HRAS^{V12}$ into the otherwise normal immortalized human breast epithelial cell line MCF10A, to obtain an early transformed cell population MCF10A-*HRAS*. MCF10A-*HRAS* was subsequently xenografted into NOD/SCID mice to develop the first-stage xenograft tumor (XT1). The cell population of the XT1 was xenografted again to develop the second-stage xenograft tumor (XT2). This procedure was repeated until two metastatic tumors were observed in the mouse carrying XT8 (Fig. S1). Cell samples from MCF10A-*HRAS*, XT1, XT2, XT3, XT4, XT5, XT6, XT7, XT8, and the two metastatic tumors XT8_M1 and XT8_M2 are in clear temporal order, together representing the cancer's full life history from initiation to metastasis. We then performed comparative genomic hybridization (Table S1), high-depth (~250X) exome sequencing (Table S2), and RNA sequencing (Table S3 and Fig. S2) for each of the cell samples, and built the first high time-resolution evolutionary roadmap of a tumor at the genomic and transcriptomic levels, revealing novel insights into cancer biology.

**Results**

**The tempo-spatial distribution of intratumoral genetic heterogeneity**

Our experimental setting enabled a tempo-spatial dissection of intratumoral genetic heterogeneity [14,15]. There are 473 single-base substitution mutations that originated after XT1 and each shows varied allele frequencies at different tumor stages (Table S2). We grouped mutations whose frequencies vary in a similar fashion, and identified five major mutation groups that together represent diverse subclones, which we validated by single cell analysis (Table S2). There is a subclone that constitutes a minor fraction of cells while harboring the vast majority of detected mutations after XT4 (Fig. 1a), suggesting it is a mutator. In this subclone we observed missense mutations on *MLL3* and *RAD54B*, two genes required for maintaining genome stability, in XT4 and subsequent stages, and a missense mutation on *PMS1*, a gene involved in DNA mismatch repair, starting in XT7 (Fig. 1b), consistent with the identification of this sbuclone as a mutator and explaining the pattern of increased mutations in this subclone (Fig. 1a). Interestingly, the two metastases XT8_M1 and XT8_M2 are seeded by the mutator subclone (Fig. 1a and Fig. S3), an observation reminiscent of a long-standing hypothesis that mutator phenotypes promote cancer evolution [16].

**The metastasis-driving positive selection on a set of multicellularity-related genes**

To understand what properties enabled the mutator subclone to seed the metastases, we studied the 137 genes that are mutated in the mutator subclone (from XT4 to XT8) using Gene Ontology (GO) analysis, and found that they are enriched exclusively in four closely related GO terms under a false discovery rate of 0.001 (Fig. 1c), while we observed no significant GO enrichment for genes mutated in other subclones. Among the four GO terms, "systems development" and "organ development" are children of the two highly similar sibling terms "multicellular organismal development" and "anatomical structure development", which share ~90% of genes with each other. Because anatomical structure is apparently a feature specific to multicelluar organisms, all these GO terms were then considered to be multicellularity-related. We examined the 15 mutated genes shared between the two sibling terms "multicellular organismal development" and "anatomical structure development", and found 14 non-synonymous (missense and nonsense) mutations and 1 synonymous mutation on them, while we observed 81 non-synonymous mutations and 43 synonymous mutations on the other 122 genes mutated in the mutator subclone, suggesting that the 15 multicelluarity-related genes tend to show functional changes ($P < 0.04$, two-tailed Fisher exact test). We ignored the single gene carrying no non-silent mutation, leaving 14 genes as a set for further examination (Table S4). None of the 14 genes belongs to the ~500 cancer drivers annotated by Cancer Gene Census (CGC) [17], and, in line with this, they are mutated only at the background frequency in breast cancer clinical samples (Fig. 1d). Interestingly, however, from a total of ~800 sequenced exomes of breast cancer clinical samples we observed 220 non-synonymous mutations and 37 synonymous mutations on the 14 genes, corresponding to an overall $d_N/d_S = 2.13$ ($P < 10^{-5}$, Binomial test; Fig. 1e and Table S4), where $d_N$ stands for the number of non-synonymous mutations per non-synonymous site, and $d_S$ stands for the number of synonymous mutations per synonymous site. This result is not due to a single outlier gene, because the $d_N/d_S$ ratio remains largely unchanged (varied from 1.85 to 2.22) after removing any one of the 14 genes, suggesting that the 14 multicellularity-related genes are overall under positive selection in breast cancer.

We noted that their tendency towards function-altering mutations appears to be stronger in our xenograft model (14:0) than in the clinical samples (220:37), although the difference is not statistically significant ($P = 0.23$, two-tailed Fisher exact test) due to limited sample sizes. We speculate that the functional alterations of the 14 genes may directly contribute to metastasis, so that the signal from the metastases-seeding subclone in our xenograft model is stronger than the signal from the clinical samples that are mostly whole primary tumors containing non-metastasizing cancer cells which dilute the signal. We examined ~100 genome-wide screenings for mutations in metastatic tumors, and discovered 82 non-synonymous mutations and 6 synonymous mutations on the 14 genes (Table S4). Thus, the tendency of having function-altering mutations is indeed stronger in metastatic tumors ($P = 0.04$, one-tailed Fisher exact test), a result well in line with our hypothesis that the positive selection on the 14 genes drives metastasis. Notably, in addition to 16 nonsense mutations, the 300 missense mutations examined here are distributed along the entire length of each of the 14 genes, without apparent hotspots (Fig. S4), suggesting that in general loss of function is positively selected for [18], although rare gain-of-function mutations may exist.

**An evolving expression profile from multicellularity to unicellularity**

We identified 12911 genes that show marked expression changes during tumor evolution. Analogous to the concept of driver mutations and passenger mutations, these changes should include both *d*river *e*xpression *c*hanges (DEC) and *p*assenger *e*xpression *c*hanges (PEC). We reasoned that, compared to DEC that are beneficial to tumor cells, PEC subject to further expression reprogramming are less likely to have fitness costs, so the PEC genes tend to show both increasing and decreasing expression during the tumor life history. With this logic in mind, we developed an algorithm and detected signature of PEC for the vast majority (~95%) of genes with marked expression changes. The remaining ~700 genes show largely one-way change (i.e., either exclusively increasing or exclusively decreasing), and thus are likely to have undergone DEC (Table S3). Consistently, gene set enrichment analysis (GSEA)[19] showed that the ~700 genes with putative DEC are involved in a large number of known cancer-related pathways/processes (Table S5). We found that up to 75.1% of these genes show reduced expressions, while the number is 38.6% for the genes with signature of PEC ($P < 10^{-16}$, Chi-square test; Fig. 2a), suggesting a "less is more" pattern in driver expression divergences. A particularly interesting observation is that, similar to the genes that are mutated in the mutator subclone (Fig. 1c), the putative DEC genes are also overrepresented in a series of multicellularity-related GO terms (Fig. 2b). Although inactivation of a specific gene can either strengthen or weaken a cellular process, the widespread shutting down of the genes necessary for development and maintenance of multicellularity suggests a general loss-of-function strategy to erase the cellular features of multicellularity during cancer.

Some of the ~700 putative DEC genes are ubiquitously expressed in diverse normal tissues and some preferentially in breast. It is interesting to know where the putative DEC genes preferentially expressed in breast evolves. Following a previous study [20], we calculated a breast biased index (BBI) for each of the ~700 putative DEC genes by dividing its expression level in breast by its median expression level in all sixteen normal tissues, and obtained 73 genes with BBI > 2 (Table S6). We clustered the expression profiles of the 73 genes in different tissues or cells. MCF10A-*HRAS*, the starting cell population, is clustered with the normal breast tissue; interestingly, the ending cell population XT8 is clustered with embryonic stem (ES) cells (Fig. 2c). In addition to their totipotency, an often underappreciated characteristic of ES cells is their rapid and excessive proliferation to form cell colony, a feature typical to unicellular life. It seems more reasonable that the unicellularity, rather than the totipotency, of ES cells is mimicked by the XT cell populations during evolution. There are 50 and 33 putative DEC genes with BBI larger than 3 and 5, respectively, with both sets showing the same pattern.

**The birth rate of cancer drivers is peaked on the deepest metazoan branch**

It is interesting to know how the findings from the experimental evolution of a breast tumor apply to clinical data of various cancers. We reasoned that, to form the genetic network required for development and maintenance of multicellularity in humans after the origin of metazoan multicellularity, some pre-existing genes are co-opted and some new genes are gradually acquired during the course of evolution. It is conceivable that genes originating at the emergence of metazoan multicellularity should have a higher chance to contribute to the core multicellularity-related genetic network. If cancer is indeed driven by demolishing the multicellularity-related genetic network, one would expect that there will be a surge in the probability of being cancer drivers for genes born at the emergence of metazoan multicellularity.

This prediction is favored by a previous phylostratigraphy analysis which shows an elevated birth rate of cancer-related human genes on the branch connecting Holozoa and Metazoa [21]. The result, however, is confounded by the then-confusing phylogeny of deep animal clades, as well as evolutionary innovations other than the emergence of multicellularity during the up to hundreds million years of evolution. A more critical issue is the methodological bias of conventional phylostratigraphy analysis which defines gene origin based on simple BLAST[22] hitting of the most distant clade/species, favoring conserved genes in deep branches. Because cancer drivers are more conserved than the rest human genes (Fig. S5), a more rigorous dating of the origin of these genes is necessary for addressing this question.

We assembled 37 completely sequenced genomes representing 13 major clades (Fig. 3a and Table S7), conducted all-against-all protein BLAST for 37*36/2 = 666 species pairs, and built an orthology network in OrthoMCL[23] that solves gene orthologous relationship using graph theory. Compared to conventional phylostratigraphy that relies on a single round of BLAST, this approach may greatly reduce the aforementioned bias. We successfully dated the origin of ~21,000 human genes, including 488 cancer drivers annotated by CGC. As a control, we carried out a simulation by randomly picking 488 genes with evolutionary rates comparable to the 488 cancer drivers, to derive the null distribution. We ran the simulation 13,000 times, and found that genes originating on the Branch #3 (or B3) are the only group comprising significantly more cancer drives than expected ($P < 0.01$ after Bonferroni correction; Fig. 3b). It is known that emergence of early metazoan clades are compressed in time [24], with an unresolved phylogenetic relationship until very recently [25]. The Branch #2 (or B2) connects unicellular life with the oldest multicellular lineage Ctenophora, so the short B3 is the deepest branch specific for metazoans (Fig. 3a). Thus, the observation of a significantly higher proportion of cancer drivers on the deepest metazoan branch is consistent with the argument that cancer is a process of demolishing the genetic network evolved for development and maintenance of multicellularity.

**The loss-of-function-dominant evolution in tumors of human patients**

There are two groups of cancer drivers, oncogenes and tumor suppressors. The former drives cancer mostly by gain of function, while the latter by loss of function. There are currently a comparable number of oncogenes and tumor suppressors characterized at CGC. Interestingly, investigation of thousands of clinical samples showed that the number of inactivated tumor suppressors often predominates over that of activated oncogenes when a single tumor is considered (the former is ~2.3 times the latter in a typical solid tumor) [18]. This number should still be an underestimation, because a large number of genes normally for maintenance of multicellularity are expected to be unrecognized minor tumor suppressors. Motivated by this prediction, we developed a new statistical test, namely the $d_T/d_S$ ratio test, to compute the relative truncating substitution mutation rate ($d_T$) to synonymous substitution mutation rate ($d_S$) in cancer; observation of $d_T/d_S$ ratio significantly larger than one indicates positive selection for null mutations, which is a unique signature of tumor suppressor genes. The design of the $d_T/d_S$ test circumvents the problem of low specificity caused by among-gene mutation heterogeneity, a major flaw of conventional methods [26], while maintaining high sensitivity in identifying minor tumor suppressors (Fig. S6). Using this new method, we analyzed data from The Cancer Genomic Atlas (TCGA) and successfully identified as many as 134 novel putative tumor suppressors under a false discovery rate of 0.1 (Table S8), a finding reminiscent of a recent study [27] and challenging the view that the growth of cancer gene list has reached a plateau [18]. It is conceivable that most of the newly identified tumor suppressors are minor cancer drivers, conferring small fitness advantages in primary tumors. Because of clonal interference[28] in which large-effect beneficial mutations suppress small-effect beneficial mutations in an asexual population, minor cancer genes may become effective mostly at late stages of cancer evolution when the pool of large-effect beneficial mutations becomes shallow, suggesting their roles on promoting malignancy or metastasis. Of particular interest is the fact that the 134 novel tumor suppressors are enriched exclusively in the same four multicellularity-associated GO terms as shown in the Fig. 1c (Fig. S7).

**Discussion**

A caveat of this study is that only a single tumor life history was examined, so some of the patterns can be rules and some are just exceptions. Importantly, the two major findings from the experimental evolution are well supported by data from clinical samples or from other studies, suggesting the generality of our results: 1) the metastasis-driving positive selection on the small set of multicellularity-related genes was first suggested by mutations in the experimental evolution, and then systematically demonstrated by data from hundreds of clinical samples; 2) the putative DEC genes were identified based on their one-way expression changes during the experimental evolution, and may contain false positives. But, this should have diluted the signal that they are overrepresented in known cancer gene sets. In addition, by using an isogenically matched cell model with three oncogenes added in an accumulative order to mimic the expression divergences during cancer, people found that ~80% of the differentially expressed genes are down-regulated [29]; analysis of these genes revealed that they are highly enriched in multicellularity-related GO terms (Fig. S8), paralleling the findings of our experimental evolution.

The idea of cancer as an evolutionary process was first coined by Nowell ~40 years ago [30]. The landmark paper distinguished cancer from genetically inheritance diseases that often have a static mutated genome, providing a basic theoretical framework to guide cancer research and cancer therapy in the past decades. Typically, gain of new genes/function seems to be the mainstream of organismal evolution [31,32], despite many reports of "less is more" [33,34]. However, cancer cells resemble unicellular life in most of their hallmarks [35], such as sustaining proliferative signaling, evading growth suppressors, resisting cell death, enabling replicative immortality, and fermentative metabolism. Although tumors are not unicellular entities as far as the physical structure is concerned, tumor cells are evolving towards the biological status in which individual cells proliferate for their private fitness, the essence of being unicellular life. Given that the whole set of cellular machinery necessary to unicellular life is already available in a human cell, cancer may reactivate the otherwise dormant pre-existing functions by knocking down the genetic constraints required for maintenance of multicellularity. Such a concept is not new, and actually quite obvious to evolutionists from a theoretical point of view [6-8], but rigorous test and systematic demonstration based on experimental/clinical data lack. In this article, we assembled various lines of empirical evidence into a reasonably convincing package for supporting the theoretical conjecture that cancer cells evolve back to be unicellular, a process reminiscent of degenerative evolution (or devolution). Equally importantly, we suggested that principles behind the cancer devolution can be very different from principles underlying typical gain-of-function-dominated organismal evolution [31]:

First, although gain of new function is sometimes necessary, for example, for initiating transformation [36], cancer devolution can be driven by the beneficial loss of function [37,38] of the existing genes that are numerous in the human genome. Considering that loss-of-function mutations are often much more accessible than gain-of-function mutations [33], devolution via loss of function seems to be a highly efficient strategy for cancer given its short-term time frame [39]. Indeed, the experimental data of this study showed that loss-of-function genetic or expression alterations are predominant, and it is also true in clinical samples [18]. Thus, the model of devolution via mainly loss of function challenges the view suggested by typical organismal evolution [31] that cancer progresses to build a new genetic network; instead, it progresses to demolish the existing multicellularity-related genetic network to reactivate the "genetic memory" of being unicellular[7]. This has important implications for thinking of the evolutionary contingency and convergence in cancer, explaining well the enormous inter-/intra-tumoral genetic heterogeneity in the clinic [40-42], because the means of knocking down a system can be numerous.

Second, the conventional evolutionary theories are also difficult to explain the origin of distant-organ metastasis within a primary tumor, because the metastatic site is often very different from the primary site in cell-biological requirements [43]. We reason that, to establish a distant-organ metastasis, cells of a primary tumor need to shut down genes that respond to signals specifying the cells' tissue identity, which is exactly the direction of devolution back to be unicellular. It is conceivable that a fully "unicellularized" cancer cell behaving like a bacterium is presumably able to colonize anywhere of the human body. Thus, distant-organ metastasis represents the very late stage of devolution which is, irrespective of external environments, internally driven by the loss of multicellularity-associated constraints. Primary tumor formation and metastatic colonization are therefore unified by the same devolutionary process. It is worth pointing out that there seems to be no major genes (like *TP53* for primary tumors) that are frequently mutated to promote metastatic colonization in clinical samples [44], suggesting that the capacity of metastasizing is a complex trait governed by a large number of mutations/genes that are either large-effect with low frequency or small-effect with high frequency, and thus calling for a quantitative genetic approach to map metastasis drivers [45].

It should be emphasized that we here argue for devolution during cancer as far as the basic features of multi-/uni-cellularity are concerned. It is apparently impossible that cancer cells degenerate to become the primitive ancestor living over 600 million years ago. While the model of devolution via loss of function offers a general theoretical framework for understanding cancer, we are fully aware that devolution is certainly not the full story of such a complex disease, and many cancer related processes, such as angiogenesis, immune evasion, tissue infiltration, and so on, may rely on evolutionary innovations rather than simple degeneration. Separation of such innovative processes from degenerative processes would be critical for designing effective cancer therapy, because attempts to stop an ongoing degeneration towards the unicellular "ground state" seem unlikely to succeed.

**Materials and Methods**
All experiments involving animals were done in the Animal Center of SYSU, in accordance to the guidelines of the center.
**The 10-stage cell populations.** A DNA fragment containing $HRAS^{V12}$ [46], an internal ribosome entry site (IRES), and the coding sequence of green fluorescent protein (GFP) was inserted into the vector *pBABE-puro*[47] to form *pBABE-puro-HRAS*$^{V12}$, the sequence of which was then verified by Sanger sequencing. This construct was introduced into the immortalized human breast cell line MCF10A that is purchased from ATCC, using a retrovirus following standard protocol. GFP positive cells were selected by flow cytometry, resulting in the MCF10A-*HRAS* cell population. About $5 \times 10^6$ MCF10A-*HRAS* cells were injected into the abdominal mammary fat pad of each of three six-week-old female NOD/SCID mice. One mouse developed a xenograft tumor about two months later, which was harvested after two months at a diameter of ~2cm and designated as XT1.

The tumor was dissected into small pieces, and suspended in pre-warmed digesting solution, prepared by adding 0.1mg/ml DNase I (Sigma DN25), 0.1mg/ml HAse (Sigma H6254), 5mg/ml Collagenase IV (Sigma C5138), 10% FBS (Gibco C20270) to DMEM-F12 (Gibco 11330-032).  After a two hour digestion at 37 ℃ the product was washed with PBS and then cultured in DMEM-F12 with 50% FBS at 37 ℃, in an incubator with 5% $CO_2$.  About $10^7$ to $10^8$ adherent cells were harvested 2 days later.  We used biotin-labeled Anti-Mouse H-2K[d] antibody (BD 553564) to label contaminating mouse cells, and subsequently depleted these with Dynabeads Biotin Binder (Life Technology 11047).  The resulting XT1 cell population was used for further xenografting, DNA and RNA extraction, and liquid nitrogen stock preparation.  The other XT cell populations were obtained through a similar procedure.

The immortalized MCF10A and MCF10A-*HRAS* cells were cultured in DMEM-F12 containing 5% Horse Serum (Gibco 16050-122), and with supplements of 20ng/ml EGF (Gibco PHG0311), 0.5mg/ml Hydrocortisone (Sigma H0888), 100ng/ml Cholera Toxin (Sigma C8052), and 10ug/ml Insulin (Sigma I1882). Because most of the above supplements are no longer required for transformed XT cells, all XT cells are cultured in DMEM-F12 containing 10% FBS, with none of the above supplements.  The change has minimal effect on gene expressions, as evidenced by the highly similar expression profiles of XT1 cell populations growing in the two media ($R = 0.98$; Table S9).

**Comparative genomic hybridization (CGH).**   Genomic DNA of ~$5 \times 10^6$ cells was extracted using DNeasy Kit (Qiagen 69504) and digested by NspI (NEB R0602L) or StyI (NEB R0500L).  Adaptors were ligated to the DNA fragments for PCR amplification.  Amplified DNA was labeled with biotin using Affymetrix Genome-Wide Human SNP Nsp/Sty Assay Kit 6.0 (Affymetrix 901015). Hybridization was performed according to Affymetrix Genome-Wide Human SNP Nsp/Sty 6.0 User Guide (Affymetrix 702504).  Arrays were then scanned by GeneChip Scanner 3000 (Affymetrix).  Genotype and copy number of each probe or genomic segment were calculated by Genotyping Console Software 4.1 (Affymetrix).  Table S1 contains the data.

**Exome sequencing.**   Genomic DNA of ~$5 \times 10^6$ cells was extracted using DNeasy Kit (Qiagen 69504), and was sheared by Covaris.  Fragments between 200bp and 300bp were collected for library construction following TruSeq DNA Guide (Illumina).  The genomic DNA library was hybridized to Human Exome 2.1M Array (Nimblegene 05-547-792-001) for exome enrichment, and then subject to Illumina GAII or HiSseq sequencing.  About 120 million 90bp reads were generated for each sample, corresponding to an average sequencing depth of ~250x for the exome.  The reads were mapped to hg19 (UCSC) by bowtie2 with default settings, and duplicated reads were removed by Picard.  Single nucleotide variants (SNVs) and indels were called on GATK platform with default settings, and only those that can pass all GATK filters were used.  The resulting data are highly reliable, as evidenced by the fact that the vast majority of variants called at a given stage were also found at the later stage. All SNVs and indels genotyped as 0/1 or 1/1 in MCF10A or MCF10A-*HRAS* were considered as germ-line SNPs or indels, and were excluded from further analysis. Table S2 contains the data.

**Poly(A)$^+$ RNA sequencing.** Total RNA of ~$3 \times 10^6$ cells was extracted using RNeasy Kit (Qiagen 74134,79654), followed by DNase I (Promega RQ1 RNase Free DNase) treatment to eliminate DNA contamination. Samples with RNA integrity number (RIN) greater than 9.5 (Agilent 2100 Bioanaylzer) were used. Poly(A)$^+$ mRNA was isolated with Dynabeads Oligo(dT)$_{25}$ (Life technology 61005). Libraries were constructed following TruSeq RNA Guide (Illumina), and subject to Illumina GAII or HiSseq sequencing. About 20 million 75bp reads were generated for each sample. The RNA-seq reads were mapped to hg19 (UCSC) by bowtie2 with default settings. Reads with mapping quality higher than 20 and mapped to exon regions (Ensemble 69) were considered as unique hits. For genes with alternative splicings, only exons from the longest transcript were considered. The RPKM of a gene was calculated similarly to previous study [48]. The effective length of a gene was defined as the total number of the 75-mers in all its exons that hits no elsewhere of the genome. We excluded genes with effective length less than 100bp from subsequent analysis. Table S3 contains the data.

**Separation of intratumoral subclones.** We started from the 8081 substitution mutations that were sequenced with >30x coverage in all samples. We excluded 7608 sites that were annotated as the same genotype by GATK in all XT samples to simplify the procedure of subclone separation. The remaining 473 sites were hierarchically clustered according to their mutant allele frequencies, which often vary at different tumor stages. We observed 72 sites that show mutations specifically in XT8_M1, and for the rest mutations we identified four major groups together representing diverse subclones. There are 15 sites that cannot be reliably grouped. The relationships of the mutation groups were resolved by reasoning and further experimental validation. To define a mutation present at a specific stage, we required that the mutant allele frequency is >1%.

We used serial dilution to obtain single cells that are seeded onto 96-well plates. Microscopic observation confirmed those wells that contain only one cell. At a very low frequency (<1%), we obtained small colonies each consisting of at least a few tens of cells in a well. Genomic DNA was extracted from such colonies and then used for assessment of intratumoral subclone separation. Table S2 contains the data.

**Determination of allele information for four mutations shared between XT8 and the two metastases.** For each of the four mutations there is a neighboring germ-line single nucleotide polymorphism (SNP) that identifies the two alleles of the locus. High-fidelity polymerase chain reactions (PCR) followed by TA cloning and Sanger sequencing were performed in XT8_M1 and XT8_M2, respectively, to trace the allele information of the mutations. Figure S3 contains the data.

**Detection of positive selection.** All mutation data from clinical cancer samples were retrieved from TCGA [49] and COSMIC [50]. The mutation spectrum of breast cancer was determined by analyzing substitution mutations in breast cancer exomes sequenced by TCGA, with only four-fold degenerate sites of protein-coding genes being considered. Using the breast cancer mutation spectrum, we then computed the number of non-synonymous (missense and nonsense) sites and synonymous sites of the merged coding sequences of the 14 multicellularity-related genes, in order to estimate their $d_N$ and $d_S$ in the breast cancer clinical samples. We examined the locations and neighboring bases for all the non-synonymous and nonsense mutations of the 14 genes observed in the xenograft tumors and the clinical samples, and confirmed that the high $d_N/d_S$ ratios cannot be explained by a few extremely biased mutation hotspots or certain biased mutation motifs. Table S4 contains the data.

**Identification of putative driver expression changes (DEC).** We defined a gene with marked expression change using two criteria: 1) there is more than two-fold difference between the maximum and minimal expressions of the 12 samples, and 2) the difference is statistically significant at $q < 0.001$ (Bonferroni correction). A total of 12911 genes show marked expression changes, with ~40% being largely down-regulated (correlation coefficient $R < 0$) during the tumor evolution. To define one-way expression divergence (or putative DEC), two rounds of analyses were carried out, each with 11 cell samples (10 earlier samples and one of the two metastases) sorted according to their temporal order. For a gene with elevated (or reduced) one-way expression divergence three criteria should be met: 1) its maximum expression among the 11 samples appears later (or earlier) than its minimal expression; 2) for all pairwise comparisons, the earlier sample cannot be higher (or lower) than the latter sample by >10% of the difference between the maximum and minimal expressions; and (3) the above two criteria are met in both rounds of analyses where a different metastasis sample is examined. Among the 12911 genes with marked expression changes, less than 6% (758) met these criteria. Fifty-five genes were completely silenced after HRAS transfection, and thus are not considered because only the stochastic evolutionary process was interested here; the remaining 703 genes are subject to further analyses. Correlation analysis revealed that ~75% of these genes were down-regulated (indicating by the negative correlation coefficient between the ranking of evolution stages and the expression levels; $R < 0$) during tumor evolution. Table S3 contains the data.

**Dating the origin of human genes.** We included 13 major clades from single-celled organisms to non-human primates, with consideration of previous reports [6,24,25,51-53] regarding the phylogeny, to trace the origin of 23,695 human protein-coding genes. Orthologous relationship of the human genes and genes of organisms in the 13 major cladenfgx s was determined using OrthoMCL with default settings. For a given human gene, we first determined the most distant clade containing the gene's orthologos, and then assigned its birth to the latest common branch of human and that clade. To avoid the potential age inflation due to horizontal gene transfer or unreliable orthology assignment, we required that the gene (or its orthologos) can be found in at least one additional clade, except for genes born at B12. Cases not satisfying this criterion were excluded, remaining 21352 human genes including 488 cancer drivers with birth time assigned. The birth rate of cancer drivers on a given branch was defined as the number of cancer drivers born on the branch divided by the total number of human genes born on the branch.

The completeness of selected genomes in the 13 clades is critical, which was evaluated using a previously described method [54]. Briefly, a set of genes basic to all eukaryotes are used to query the selected genomes, and observation of 100% presence is desired. We selected 36 species with sequenced genomes to represent the 13 clades, with a minimal completeness of ~96% at the clade Placozoa. Inclusion of more genomes does not improve the performance. The major point of this analysis lies in the three clades (Placozoa, Porifera, and Ctenophora) representing early multicellular animals. To model the potential bias due to sparse sampling of the three clades, we performed 100 times simulations by randomly removing 10% of genes of the genomes in each of the three clades, and found that the overrepresentation of cancer genes at the Branch #3 remains qualitatively unchanged (Fig. S9).

To control for protein evolutionary rate in the phylostratigraphy analysis, we sorted the 21352 genes according to their $d_N$ between human and chimpanzee (the results are essentially the same when $d_N$ between human and rhesus monkey were used.), and then divided them into 100 equal-size bins (the fastest evolving bin has 213+52=265 genes). We randomly pick $n$ genes from a bin, where $n$ is the number of cancer drivers in the bin, to form a pseudo-cancer driver set with 488 genes; the birth rate of the pseudo-cancer drivers on each of the 13 branches is then calculated. Such simulation was repeated 13,000 times, to make a test at the significance level of $P = 0.01$ after Bonferroni correction. Table S7 contains species and gene information used in this analysis.

**Indentification of putative tumor suppressor genes using $d_T/d_S$ ratio test.** We downloaded TCGA level 3 data of substitution mutations from TCGA data portal; in total, 776 BRCA, 269 COAD, 291 GBM, 306 HNSC, 422 KIRC, 520 LUAD, 178 LUSC, 463 OV, 116 READ, 266 SKCM, 245 STAD, 406 THCA and 248 UCEC tumors were considered. For each cancer type we analyzed the single-base substitution mutations at four-fold degenerate sites of protein-coding genes to derive mutation spectrum of that cancer. In addition to the six regular types of substitution mutations, mutations of ApT->ApA, CpG->TpG, TpC->TpX (X: A, T, and G), and TpCpG->TpTpG (the combination of TpC and CpG) were separately considered [26]. In other words, we computed the mutation rate $U$ for each of the 12 mutation types in each cancer to estimate the expected *t*runcating vs *s*ynonymous (T/S) site ratio of a gene. For a hypothetical gene with 100 codons, we would consider 900 mutation possibilities; for each mutation possibility the expected rate is the $U$ of the corresponding mutation type. We then sum up the expected rates (i.e., all $U$) of all truncating mutation possibilities and all synonymous mutation possibilities, and the expected T/S site ratio is the former divided by the latter. The $d_T/d_S$ ratio is the observed T/S mutation ratio divided by the expected T/S site ratio. Multiple testing was controlled using the method of Storey [55]. Table S8 contains the data.

**Software or URLs.** Bowtie2 [56] for short reads mapping; Picard (http://picard.sourceforge.net) for removing duplicated short reads; GATK [57] *for SNVs and indels calling; Cluster 3.0* [58] *for clustering analysis; R-package affy* [59] for microarray data processing; BinGO [60] and GSEA for Gene Ontology analysis; ENSEMBL http://asia.ensembl.org; UCSC hg19 http://genome.ucsc.edu; ENCODE http://genome.ucsc.edu/ENCODE/; Bodymap 2.0 http://www.ncbi.nlm.nih.gov/geo/query/acc.cgi?acc=GSE30611; GO http://www.geneontology.org/;TumorScape http://www.broadinstitute.org/tumorscape; OrthoMCL http://orthomcl.org/orthomcl/.

**Supplementary Information** can be found with this article online.

**Acknowledgements**
We thank Professors J. Zhang (U. of Michigan), Z. Songyang (Baylor College of Medicine), H. Liang (M.D. Anderson Cancer Center), and X. Liu (Harvard) for comments and critical reading of the manuscript. This work was supported by NSFC.

**Figure Legends**

**Figure 1. The evolving intratumoral genetic heterogeneity reveals positive selection on a set of multicellularity-related genes. a.** The area of a circle is proportional to the cell frequency of the subclone, which is also shown next to the circle (mean±s.e.m.). The unfilled circle represents the subclone that became a mutator after XT4. The number of mutations observed in the mutator subclone and in remaining parts of the tumor is shown at the left. **b.** Allele frequency of the three missense mutations on *MLL3*, *RAD54B*, and *PMS1*. **c.** GO terms enriched with genes mutated in the mutator subclone. Arrows stand for "is_a", and FDR is false discovery rate. The number of mutated genes in each GO term is shown inside each circle. **d.** The rate of non-silent mutations (missense, nonsense, and splicing-altering substitutions as well as indels) for each of the 14 multicellularity-related genes with non-synonymous mutations in the mutator subclone, and for all human protein-coding genes in a total of 772 breast cancer samples in TCGA. **e.** $d_N > d_S$, a signature of positive selection on the 14 genes in breast cancer clinical samples. Error bar shows one standard error.

**Figure 2. Functional properties of the ~700 genes with putative DEC. a.** The percentage of reduced expressions for putative DEC (*d*river *e*xpression *c*hange) genes and PEC (*p*assenger *e*xpression *c*hange) genes, respectively. Error bar shows one standard error. **b.** The putative DEC genes are enriched in multicellularity-related GO terms. Arrows stand for "is_a", and FDR is false discovery rate. **c.** An evolving expression profile approaching that of embryonic stem cells.

**Figure 3. The birth rate of cancer drivers is peaked on B3, the deepest branch of metazoans. a.** The phylogenetic relationship of human and the 13 major clades. Emergence of the four early metazoan clades Ctenophora, Porifera, Placozoa, and Cnidaria was extremely compressed in time ~600 million years ago. The number of born genes is shown in parenthesis behind each branch, and the number of used genomes is shown behind each clade. **b.** The birth rate of cancer drivers and pseudo-cancer drivers (y-axis), randomly selected genes with evolutionary rates comparable to cancer drivers, on each of the 13 branches (x-axis). Red triangles are for cancer drivers, and box plots are for pseudo-cancer drivers, with horizontal lines showing the median rates and black dots showing the $10^{th}$ or the $12990^{th}$ rates of a branch out of 13,000 simulations.

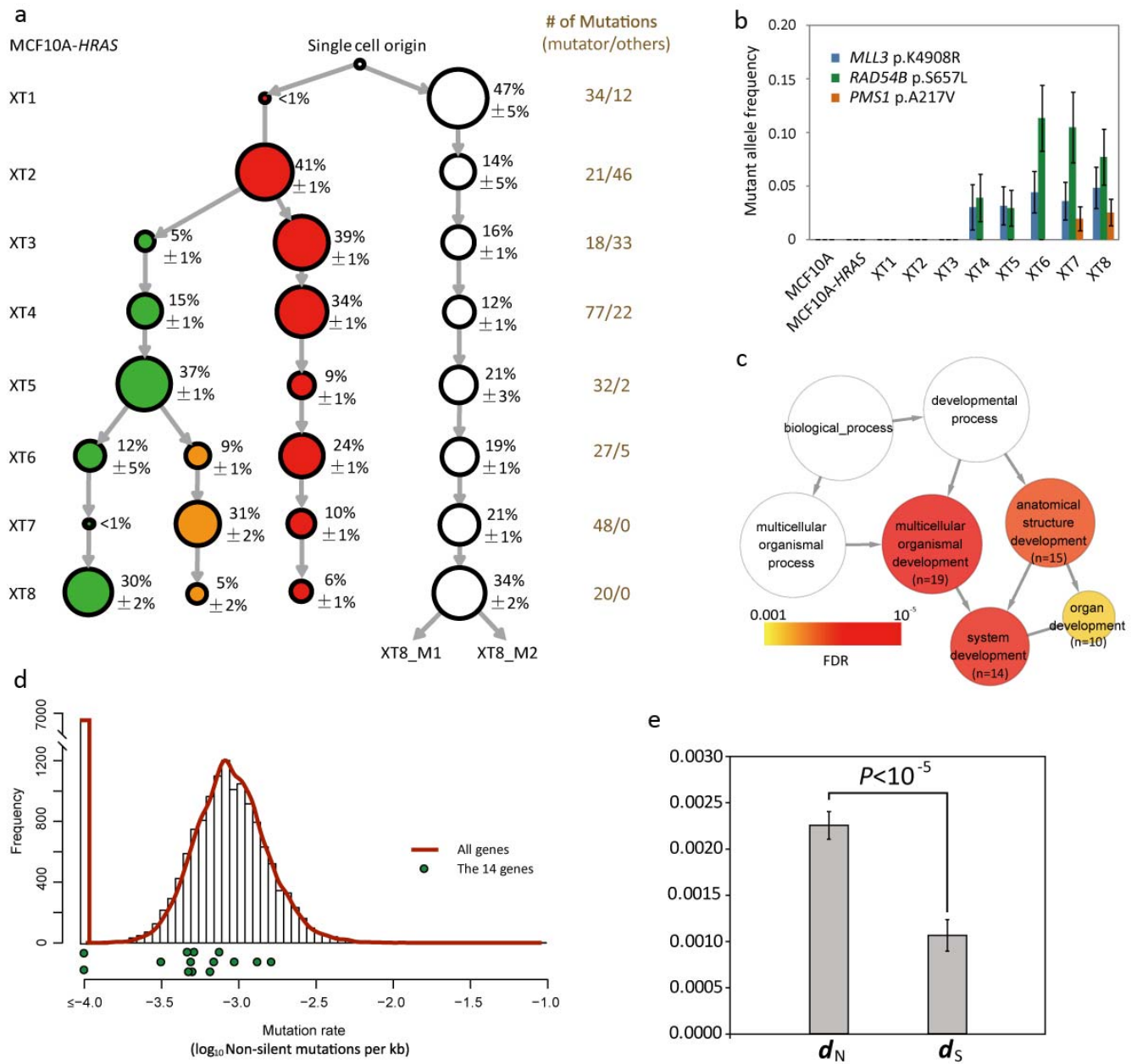

**Figure 1**

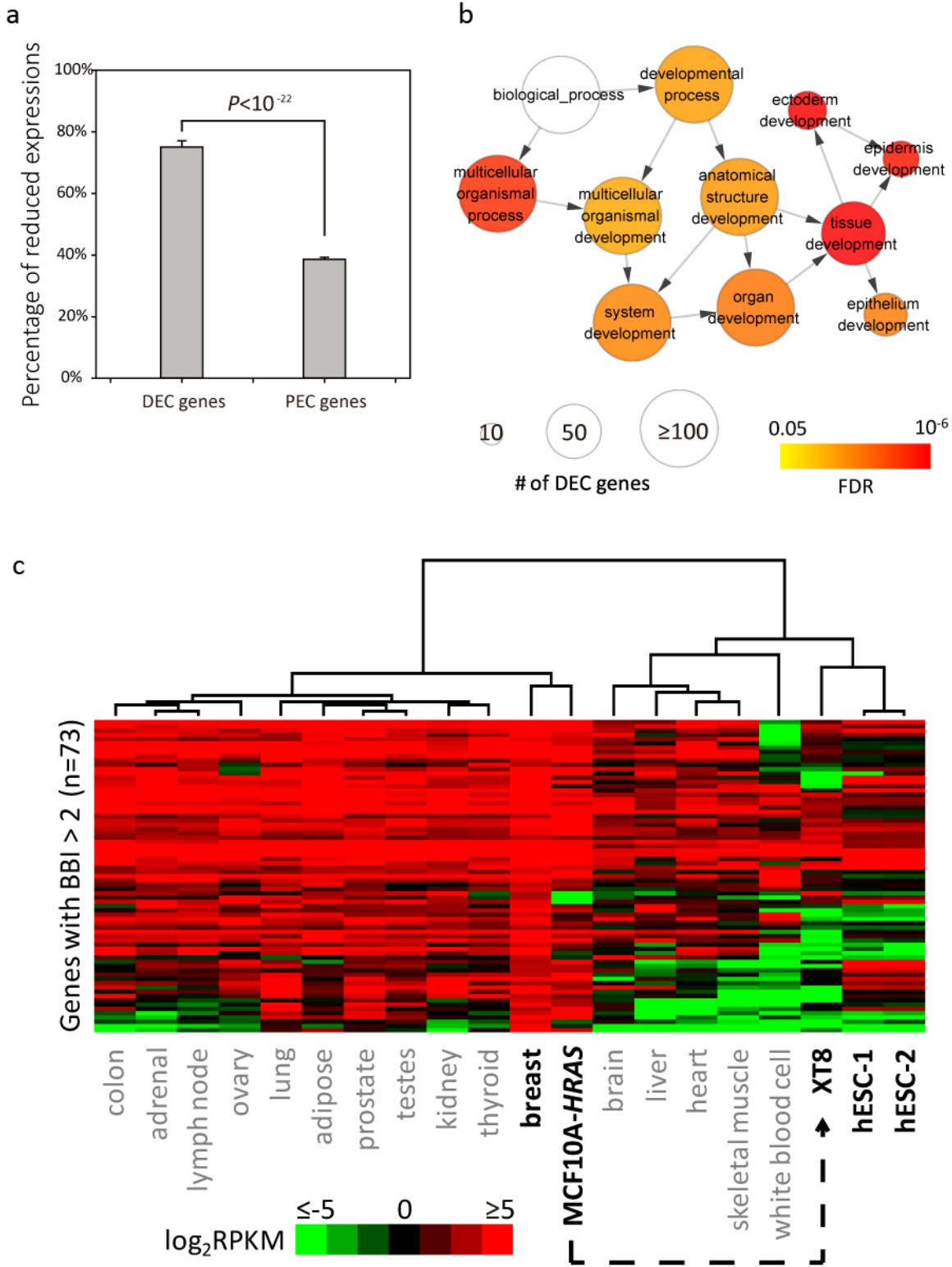

**Figure 2**

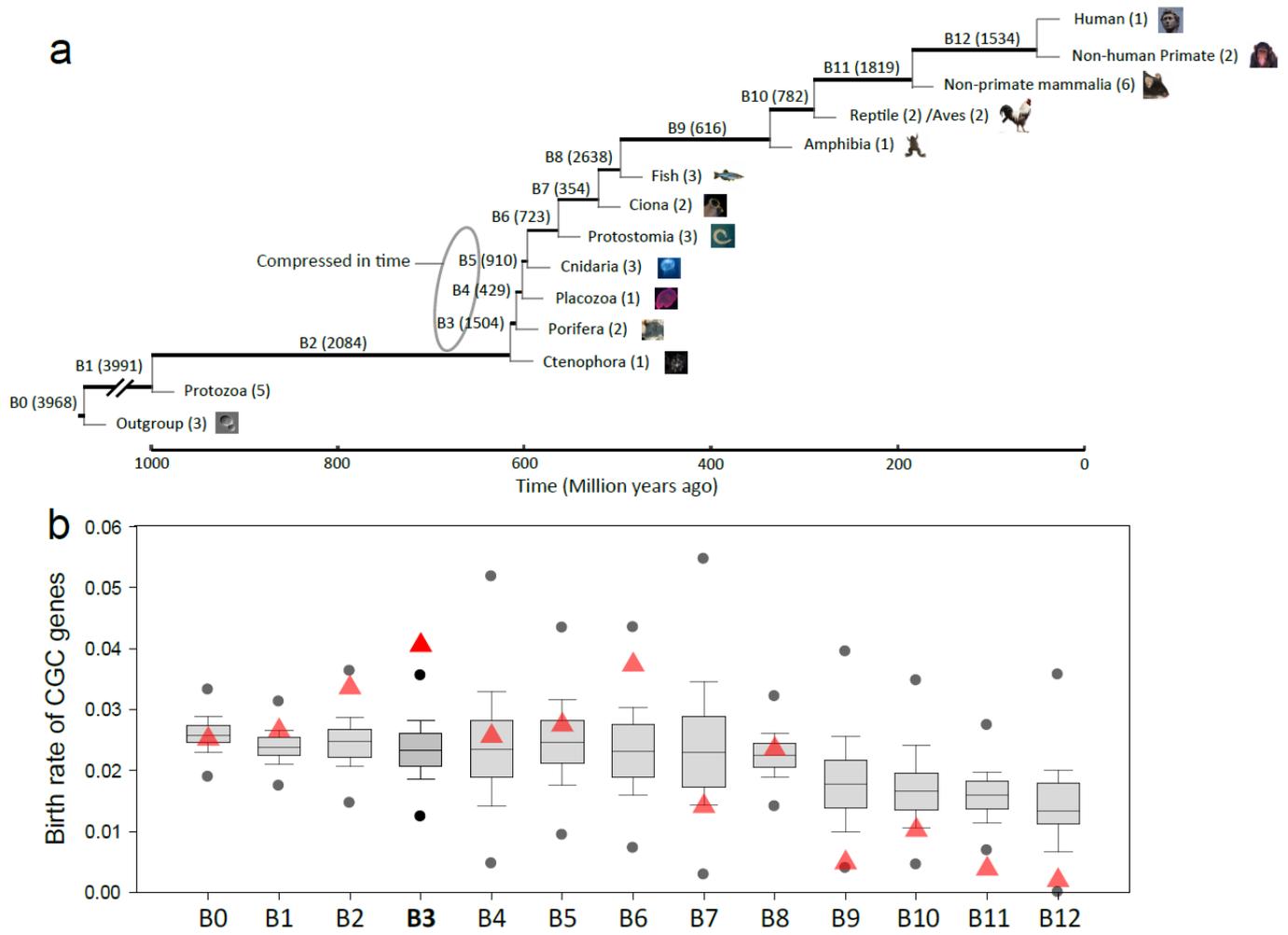

**Figure 3**

Supplementary Information of "The degenerative evolution from multicellularity to unicellularity during cancer"

Han Chen, Fangqin Lin and Xionglei He

State Key Laboratory of Biocontrol, School of Life Sciences, Sun Yat-sen University, Guangzhou, 510275, China

**Supplementary Information includes:**
Legends for Supplementary Tables 1-9 (Supplemental Tables are provided in separated files)
Legends for Supplementary Figures 1-9
Supplementary Figures 1-9

**Legends for Supplemental Tables 1-9**

**Table S1    Copy number information during the tumor evolution.** This table contains copy number information of 10356 genomic regions in the 12 samples identified by comparative genomic hybridization (CGH).    There are 13 regions that show one-way copy number change during the tumor progression, and encompass breast cancer drivers identified by TumorScape.    It also includes 24 cancer driver genes annotated in Cancer Gene Census that are affected by copy-number-neutral loss of heterozygosity.

**Table S2    DNA alterations revealed by exome sequencing.**    This table contains all single nucleotide variants (or mutations) and small indels covered by at least 30 mapped reads in each of the 12 samples.    It also contains information of the 473 sites that were mutated after XT1 and considered in subclone separation.    In addition, the single-cell sequencing results that were used to assess the subclone separation are also included (grey represents mutations that are not grouped).

**Table S3    Expressions measured by log$_2$RPKM for all the human genes annotated in Ensembl.**    It also contains information of DEC and PEC genes.

**Table S4    Raw data for computing $d_N$ and $d_S$ of the 14 multicellularity-related genes.**

**Table S5    Cancer gene sets enriched with the ~700 putative DEC genes.**

**Table S6    The log$_2$RPKM of the 73 breast biased genes in MCF10A-*HRAS*, XT8, 16 normal tissues and two human embryonic stem cells (hESCs).**    We obtained RNA-seq data of the 16 normal tissues from Bodymap 2.0, and data of the two hESCs from ENCODE.

**Table S7    Genes and genomes in the phylostratigraphy analysis.**

**Table S8    Tumour suppressor genes revealed by the $d_T/d_S$ ratio test.**    For each cancer type, the $d_T/d_S$ ratio and the corresponding $p$-value (binomial test) are computed for each of the genes with ≥10 single-base substitution mutations in their coding sequences.    Genes with $d_T/d_S$ significantly larger than 1, under the cutoff of $q$ < 0.1 after controlling for multiple testing, are shown.

**Table S9    Gene expressions (log$_2$RPKM) of XT1 cells growing in media with and without supplements.**

**Legends for Supplemental Figures 1-9**
**Fig. S1  A schematic diagram of the xenograft experimental evolution.**

**Fig. S2  The phylogenetic relationship of the tumors revealed by their expression distance.**   The expression distance of two samples was measured using 1-*R*, where R is the Pearson correlation coefficient of two $\log_2$-expression profiles, and the FM method (PHYLIP 3.69) was used to build the distance-based tree.

**Fig. S3 Determination of allele information for four mutations shared between XT8 and the two metastases.**   The mutant allele frequencies in XT8 are 15.4%, 4.8%, 5.3%, and 9.1% for locus 1, 2, 3, and 4, respectively.   TA-cloning based Sanger sequencing indicated that the G->A mutation is 100% present at one allele of locus 1 in the two metastases, suggesting that the two metastases originated exclusively from an XT8 subclone carrying this mutation.   Meanwhile, each of the three mutations at loci 2-4 is not 100% present at a specific allele, suggesting multiple cell origin of the two metastases.

**Fig. S4  The distribution of missense, nonsense, and synonymous mutations on the coding sequence of each of the 14 multicellularity-related genes.**   The relative positions of the mutations on the coding sequences are shown.

**Fig. S5  Comparison of protein length and protein evolutionary rate between cancer drivers and the rest human genes.   a.** The encoded protein length is similar between cancer drivers and the rest genes (K-S test is used).   **b.** Cancer drivers evolve slower than the rest genes (K-S test is used).

**Fig. S6  Comparison of the $d_N/d_S$ test and $d_T/d_S$ test in predicting ten canonical TSGs.**   Pooled mutations from 4506 tumor samples of 13 cancer types were analyzed.

**Fig. S7 The novel tumour suppressors are enriched exclusively in the same four multicellularity-related GO terms as in the Fig. 1c.**   The number of novel tumour suppressors in each GO term is shown in parenthesis.   FDR stands for false discovery rate.

**Fig. S8  The genes differentially expressed in the four-stage cell model are highly enriched in multicellularity-related GO terms.**   The number of differentially expressed genes in each GO term is shown in parenthesis.   FDR stands for false discovery rate.

**Fig. S9 The observation of more cancer drivers on B3 holds after randomly removing 10% of genes from each of the three early animal clades, Placozoa, Porifera, and Ctenophora.**   For Ctenophora and Placozoa, 10% of genes were randomly dropped out from the *Trichoplax adhaerens* and *Mnemiopsis leidyi* genomes,

respectively. For Porifera, 10 % of genes were removed from both the *Amphimedon queenslandica* and the *Oscarella carmela* genomes. The simulations were repeated 100 times for each clade.

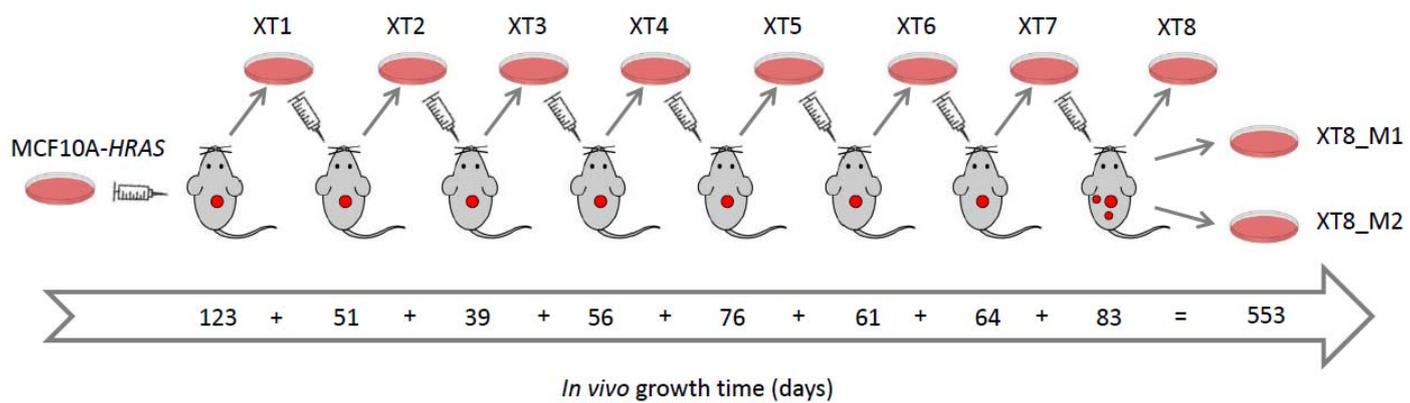

**Fig. S1**

**Fig. S2**

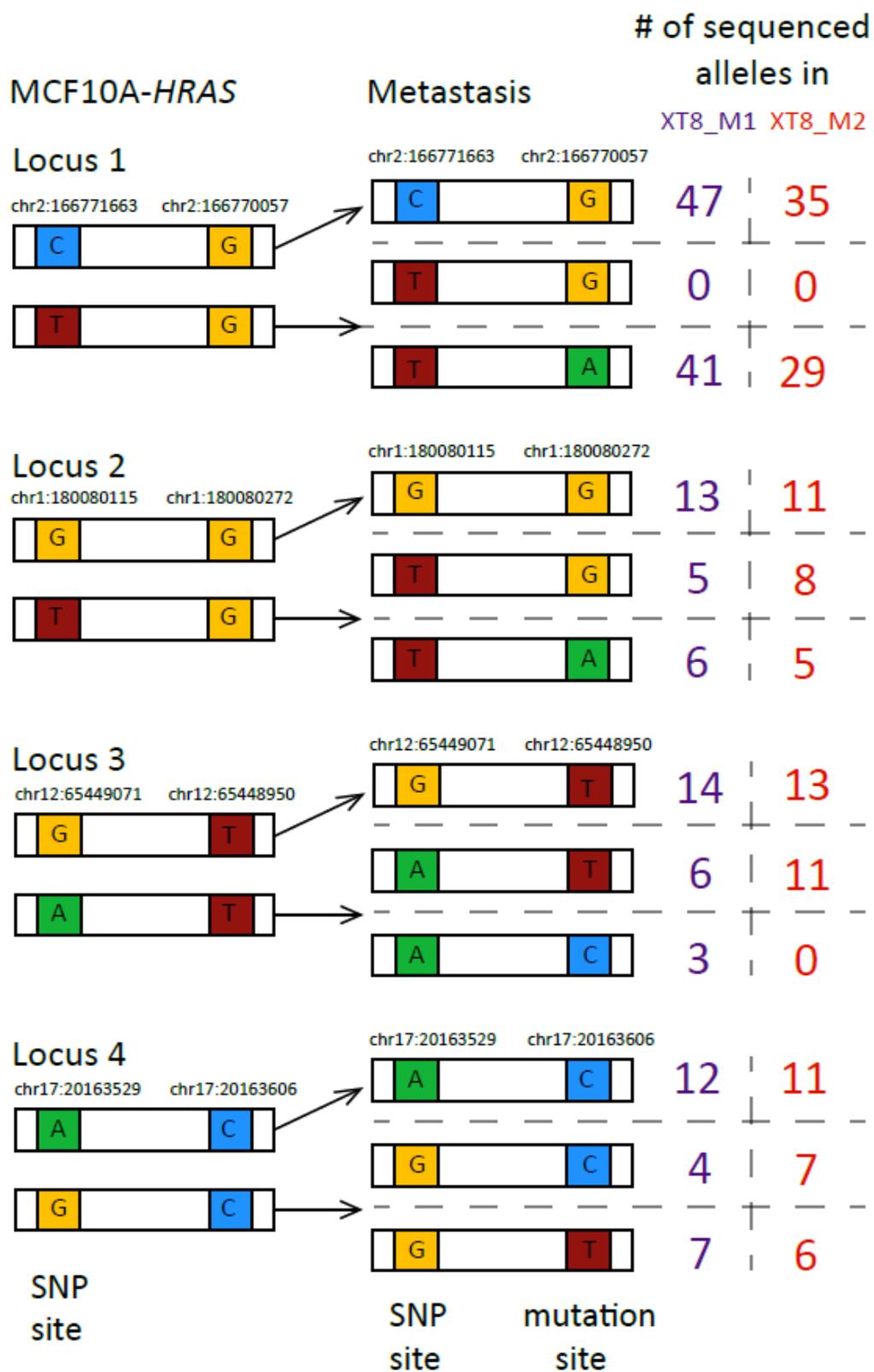

**Fig. S3**

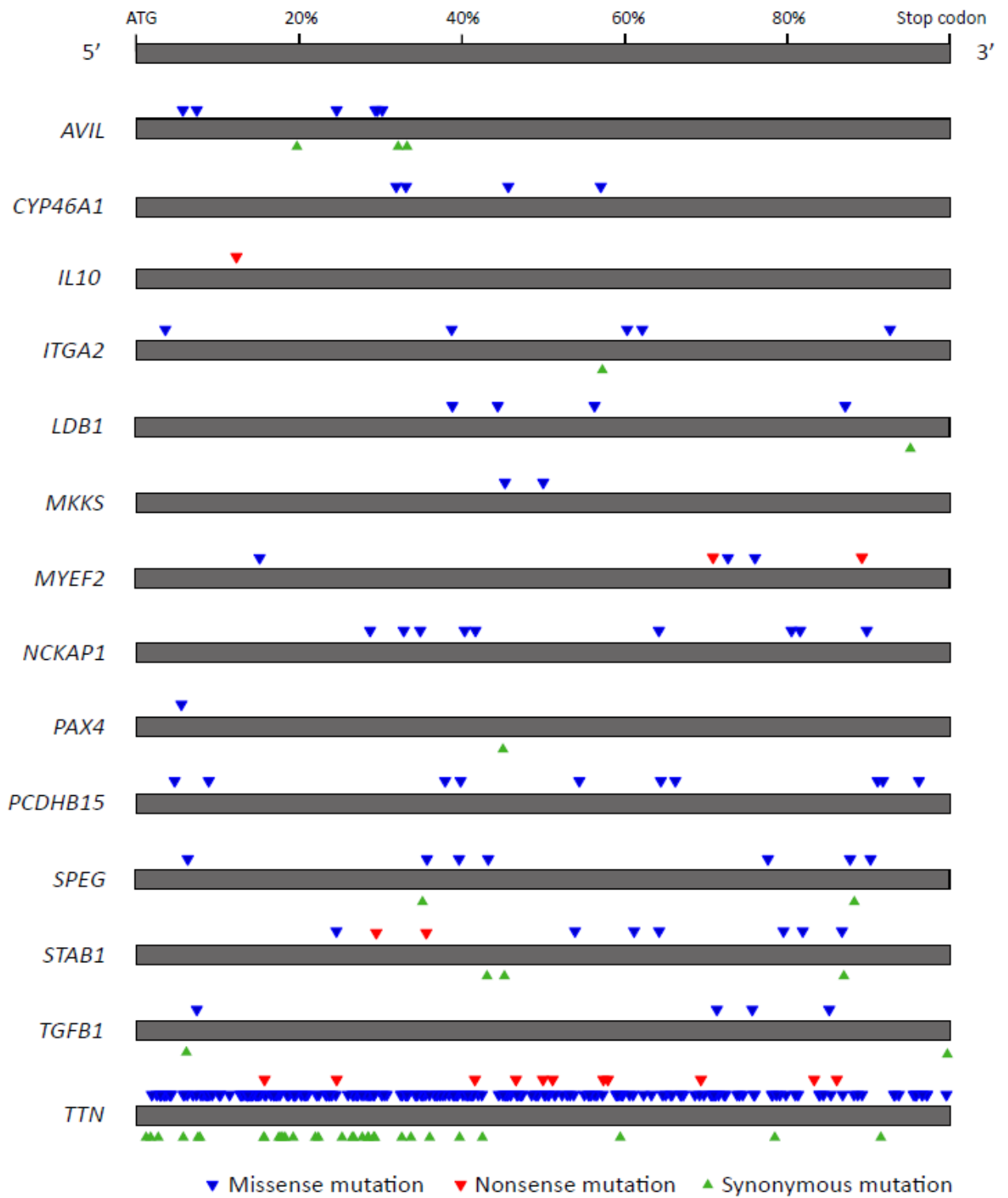

**Fig. S4**

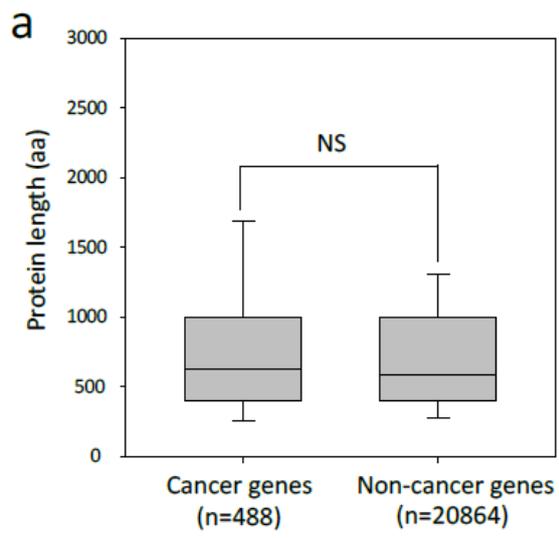 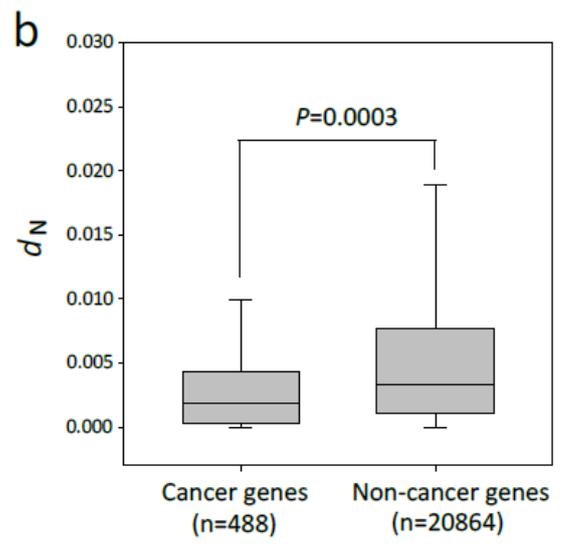

**Fig. S5**

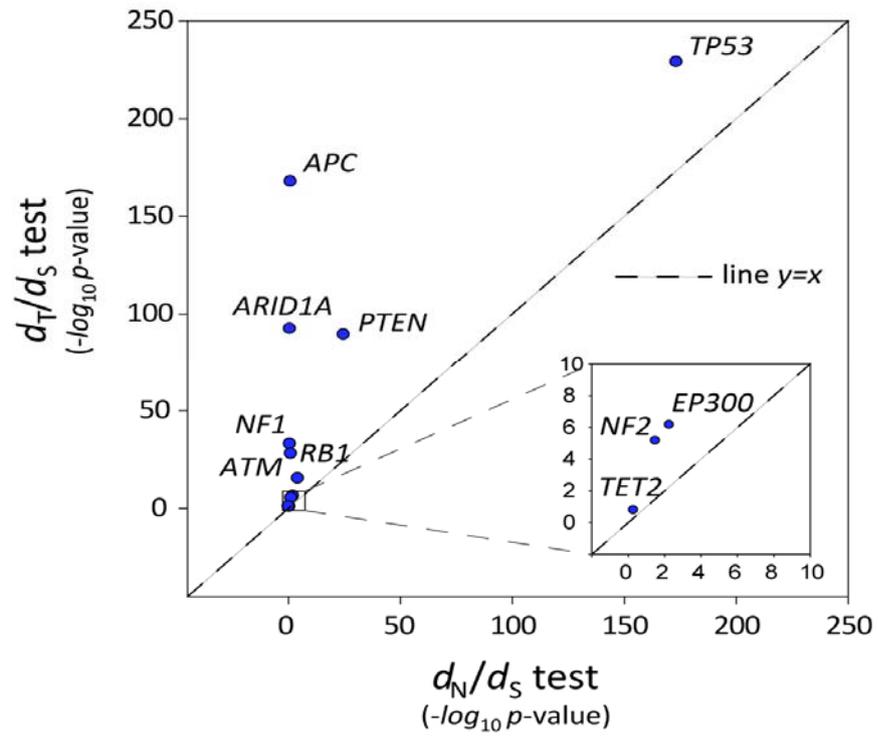

**Fig. S6**

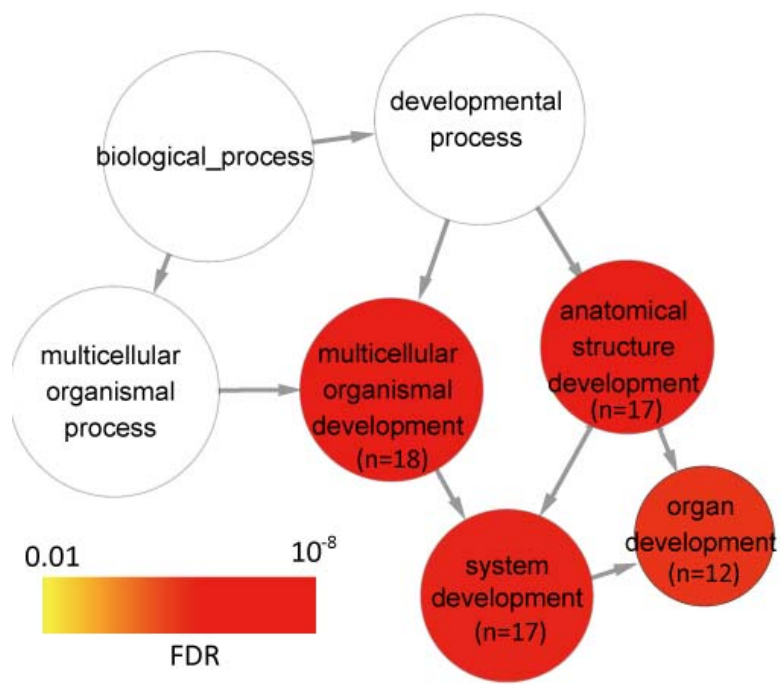

**Fig. S7**

**Fig. S8**

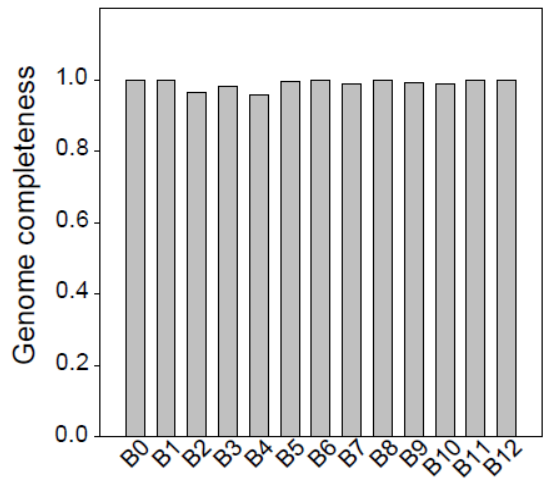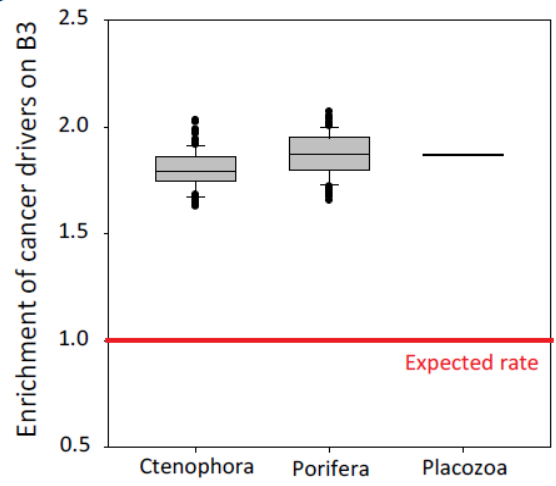

**Fig. S9**